\documentclass[twocolumn,aps,superscriptaddress,showpacs]{revtex4}
\usepackage{amssymb}
\usepackage{amsmath}
\usepackage{mathrsfs}
\usepackage{graphicx}
\usepackage{subfigure}
\usepackage[normalem]{ulem}
\usepackage[dvips]{color}

\begin{document}

\title{Symmetry energy effects on the properties of hybrid stars}
\author{He Liu}
\email{liuhe@qut.edu.cn}
\affiliation{Science School, Qingdao University of Technology, Qingdao 266000, China}
\affiliation{The Research Center of Theoretical Physics, Qingdao University of Technology, Qingdao 266033, China}
\author{Jun Xu}
\email{xujun@zjlab.org.cn}
\affiliation{Shanghai Advanced Research Institute, Chinese Academy of Sciences,
Shanghai 201210, China}
\affiliation{Shanghai Institute of Applied Physics, Chinese Academy of Sciences, Shanghai 201800, China}
\author{Peng-Cheng Chu}
\email{kyois@126.com}
\affiliation{Science School, Qingdao University of Technology, Qingdao 266000, China}
\affiliation{The Research Center of Theoretical Physics, Qingdao University of Technology, Qingdao 266033, China}
\date{\today}
\begin{abstract}
Symmetry energy is an important part of the equation of state of isospin asymmetry matter. However, the huge uncertainties of symmetry energy remain at suprasaturation densities, where the phase transitions of strongly interacting matter and the quark matter symmetry energy are likely to be taken into account. In this work, we investigate the properties of symmetry energy by using a hybrid star with the hadron-quark phase transition. The interaction among strange quark matter (SQM) in hybrid stars is based on a 3-flavor NJL model with different vector and isovector channels, while the equation of state (EOS) of the nuclear matter is obtained by considering the ImMDI-ST interaction by varying the parameters $x$, $y$, and $z$. Our results indicate that the various parameters and coupling constants of the interactions from the ImMDI-ST and NJL model can lead to widely different trends for the symmetry energy in the hadron-quark mixed phase and the different onsets of the hadron-quark phase transition. In addition, it has been found that the radii and tidal deformabilities of hybrid stars constrain mostly the density dependence of symmetry energy while the observed maximum masses of hybrid stars constrain mostly the EOS of symmetric nuclear and quark matter.

\end{abstract}

\pacs{21.65.-f, %Nuclear matter
      21.30.Fe, %Forces in hadronic systems and effective interactions
      51.20.+d  %Viscosity, diffusion, and thermal conductivity
      }

\maketitle
\section{Introduction}
\label{INT}
Studying the nature and constraint the equation of state (EOS) of isospin asymmetric matter is one of the main scientific goals of nuclear physics. Researches based on terrestrial nuclear experiments~\cite{Dan02,Bar05,BLi08,Bor19,Ono19,Ste05} and astrophysical observations of compact stars~\cite{Lat00,Cxu10,Che12,Wat16,Oer17,Bom18} have already led to significant constraints on the EOS of symmetric nuclear matter (SNM) and the symmetry energy $E_{sym}(\rho)$ around but mostly below the saturation density of nuclear matter ($\rho_0 \approx 0.16$ fm$^{-3}$). However, our knowledge on the symmetry energy at supra-saturation densities, where the phase transitions of strongly interacting matter and quark matter symmetry energy are likely to be taken into account, is still limited. The EOS of nuclear matter at densities $2\rho_0 < \rho < 5\rho_0$ has also been constrained by the measurements of collective flows~\cite{Dan02} and subthreshold kaon production~\cite{Fuc06} in relativistic heavy-ion collisions. As pointed out in Refs. ~\cite{Dan02,Pie04,Fuc06}, remaining uncertainties in the determination of the EOS of nuclear matter are mainly related to those in density dependence of symmetry energy. In fact, the high-density $E_{sym}(\rho)$ has been broadly recognized as the most uncertain part of the EOS of isospin asymmetric matter~\cite{Bal16,BLi17,BLi18}.

Compact stars are the natural testing ground of the isospin-dependence of strong interactions and the corresponding EOS of isospin asymmetric matter at high densities and large isospin asymmetries. The radii of compact stars are known to be determined by the pressure at densities around
$2\rho_0$~\cite{Lat00,Lat01}, which are thus sensitive to the density dependence of symmetry energy in this density region. In recent reports, a radius measurement based on Neutron Star Interior Composition Explorer (NICER) and X-ray Multi-Mirror (XMM-Newton) found that the radius of PSR J0740+6620 is $13.7^{+2.6}_{-1.5}$ km ($68\%$ credibility)~\cite{Mil21}. In particular, the radius range that spans the $\pm1\sigma$ credible intervals of all the radius estimates in the different frameworks is $12.45\pm 0.65$ km for a $1.4M_{\odot}$ compact star~\cite{Mil21}. The x-ray bursts from accreting neutron stars in low-mass x-ray binary systems also provide potential possibilities to constrain the mass and radius simultaneously~\cite{Lat14,Mil16,Sul16}. In Ref.~\cite{Lat14}, the radius of compact stars with the canonical mass $M=1.4M_{\odot}$ has been constrained to the range of $10.62 \leq R_{1.4} \leq 12.83$ km. The mass of compact stars is also the main astrophysical observable that can be used to extract information on the EOS of strongly interacting matter~\cite{Lat07}. The measurement of PSR J1614-2230 and PSR J0348+0432 a few years ago had led to a precise determination of $1.97\pm0.04M_{\odot}$~\cite{Dem10} and $2.01\pm0.04M_{\odot}$ for their respective masses~\cite{Ant13}, while the newly measured gravitational mass of PSR J0740+6620 is declared as $2.08\pm0.07M_{\odot}$, which is considered as the  highest reliably determined compact star mass~\cite{Cro20}. More recently, the gravitational wave events GW170817~\cite{Abb17} and GW190814~\cite{Abb20} have provided more additional constraints on the EOS of the compact star matter. The analysis of GW170817 by the LIGO/Virgo Collaboration has found with a $90\%$ confidence that the tidal deformability of the merging neutron stars constrained as the range $70 <\Lambda_{1.4}<580$~\cite{Abb18}. Many studies have used the measurement of the tidal deformability from GW170817 to derive new constraints on the nuclear symmetry energy~\cite{Kra19,Car19,Zha19}. Moreover, the newly discovered compact binary merger GW190814~\cite{Abb20} which has a secondary component of mass $(2.50 \sim 2.67) M_{\odot}$ at $90\%$ credible level has also aroused lots of debates on whether the candidate for the secondary component is a compact star or a light black hole. The existence of such high-mass compact stars indicates that the EOS of compact star matter is relatively stiff and it yields high pressures at a few times saturation density $\rho_0$. In the present study, we will employ the above constraints on the compact star mass, radius, and tidal deformability to extract information on the properties of isospin asymmetric matter in compact stars.

Although the EOS of a pure nucleonic matter can generally be stiff enough to support a two-solar-mass compact star, hyperon as well as quark degrees of freedom are expected to appear with the increasing baryon chemical potential, forming the so-called hybrid stars. It has been found that the high-mass constraint may be used to understand the properties of the hadron-quark phase transition as well as the EOS of the mixed phase in hybrid stars (see, e.g., Refs.~\cite{Sho03,Alf13,Bay16,Han19,Alv19}). Also, the radius of compact stars has been shown to be closely related to the isovector part of the EOS of quark matter~\cite{BLi06}. In the present study, we consider a hybrid star with a quark core at high densities, a hadron-quark phase transition (mixed phase) at moderate densities and a hadronic phase at low densities. The hadron-quark phase transition is one of the most concerned topics, and some recent evidences indicate quark-matter cores can appear in massive compact stars~\cite{Ann20}. Moreover, it is also an important topic to further explore the QCD phase structure and search for the signal of the critical point in heavy-ion collisions. The hybrid star with quark matter in the inner core seems to have problems in describing massive compact stars, which is due to a lack of sufficient repulsion of the quark matter effective interactions leading to a soft EOS of quark phase at high densities. To solve this problem, we investigate the properties of quark matter based on the 3-flavor NJL model with vector and isovector couplings. Both the vector and isovector interactions have been shown to have important impacts on the quark matter EOS and the QCD phase structure~\cite{Asa89,Fuk08,Bra13,Chu14,Chu15,Chu16,Chu19,Liu16}. Meanwhile, we describe nuclear matter using an improved isospin- and momentum-dependent interaction (ImMDI) model, which is constructed from fitting cold nuclear matter properties at saturation density and the empirical nucleon optical potential~\cite{Jxu15}. And it has been extensively used in intermediate energy heavy-ion reactions to study the properties of asymmetric nuclear matter. The ImMDI interaction, in the present paper, will be modified and used to study the equation of state of nuclear matter and the properties of hybrid stars by including the phase transition from nuclear matter to quark matter.

\section{The theoretical model }
\label{MODEL}
Here we apply the 3-flavor NJL model and ImMDI model to hybrid stars, with a quark core at high densities, a mixed phase of quarks and hadrons at moderate densities, and a hadronic phase at low densities. The possible appearance of hyperons is neglected, which is due to the fact that there are still large uncertainties on the hyperon-nucleon (\emph{YN}) and hyperon-hyperon (\emph{YY}) interactions in the nuclear medium~\cite{Hiy20,Con16}. Besides, the presence of new degrees of freedom, such as hyperons, tends to soften the equation of state at high densities and lower the maximum mass of compact stars~\cite{Yan08}. Furthermore, following the results from Ref.~\cite{Jxu10}, the EOS of hybrid star matter is mostly dominated by the hadron-quark phase transition because the fraction of hyperons disappears quickly in mixed hadron-quark phase, which means that the effect of hyperons on the symmetry energy, especially in the hadron-quark mixed phase, is expected to be small. Thus, in this work we mainly focus on symmetry energy effects on the properties of hybrid stars without hyperons.

In the high-density quark phase, the system consists of a mixture of quarks ($u$, $d$, and $s$) and leptons ($e$ and $\mu$) at charge neutrality
\begin{equation}
\frac{2}{3}\rho_u-\frac{1}{3}(\rho_d+\rho_s)-\rho_e-\rho_\mu=0,
\end{equation}
and the $\beta$-equilibrium condition in quark phase. It is given by $\mu_i=\mu_Bb_i-\mu_cq_i$ with $\mu_B$ and $\mu_c$ being the baryon and charge chemical potentials of quark phase, respectively. $q_i$ and $b_i$ are, respectively, charge and baryon numbers of the particles. The detailed $\beta$-equilibrium conditions are given by
\begin{eqnarray}
\mu_s&=&\mu_d=\mu_u+\mu_e,
\\
\mu_\mu&=&\mu_e.
\end{eqnarray}
For quark matter, the energy density ($\varepsilon_Q$) and the pressure($P_Q$) can be obtained from the NJL model. In the mean-field approximation, the energy density $\epsilon_Q$ of quark matter from the NJL model with vector and isovector couplings in detail can be written as~\cite{Liu16}
\begin{eqnarray}
\varepsilon_Q &=&-2N_c\sum_{i=u,d,s}\int_0^\Lambda\frac{d^3p}{(2\pi)^3}
E_i(1-f_i-\bar{f}_i)
\notag\\
&-&\sum_{i=u,d,s}(\tilde{\mu}_i-\mu_i)\rho_i+G_S(\sigma_u^2+\sigma_d^2+\sigma_s^2)
\notag\\
&-&4K\sigma_u\sigma_d\sigma_s-G_V(\rho_u^2+\rho_d^2+\rho_s^2)
\notag\\
&+&G_{IS}(\sigma_u-\sigma_d)^2-G_{IV}(\rho_u-\rho_d)^2-\varepsilon_0.
\end{eqnarray}
In the above, the factor $N_c=3$ represents the color degeneracy of quark, as well as $f_i$ and $\bar{f}_i$ are respectively the Fermi distribution functions of quark and antiquark with flavor $i$. $\sigma_i$ and $\rho_i$ stand for the quark condensate and the net quark number density, respectively; $\tilde{\mu_i}$ is the effective chemical potential which depends on the vector and isovector interactions~\cite{Liu16,Liu20}; $E_i(p)=\sqrt{p^2+M_i^2}$ is the single quark energy; and $\varepsilon_0$ is introduced to ensure $\varepsilon_Q=0$ in vacuum. $G_S$ and $G_V$ are the strength of the scalar and vector coupling, respectively; and the $K$ term represents the six-point Kobayashi-Maskawa-t'Hooft (KMT) interaction that breaks the axial $U(1)_A$ symmetry~\cite{Hoo76}. The additional $G_{IS}$ and $G_{IV}$ terms represent the scalar-isovector and the vector-isovector interactions, respectively. For the ease of discussions, we define the relative strength of the vector coupling, the scalar-isovector coupling and the vector-isovector coupling respectively as $R_V=G_V/G_S$, $R_{IS}=G_{IS}/G_S$ and $R_{IV}=G_{IV}/G_S$. As is known, the position of the critical point for the chiral phase transition is sensitive to $R_V$~\cite{Asa89,Fuk08,Bra13}, which was later constrained within $0.5 <R_V <1.1$ from the relative $v_2$ splitting between protons and antiprotons as well as between $K^+$ and $K^-$ in relativistic heavy-ion collisions~\cite{JXu14}. Also, the strong vector-isovector interaction seems to be needed to reproduce the $v_2$ difference between $\pi^+$ and $\pi^-$ with the same NJL transport approach at the same collision energies~\cite{Liu19}. The strength of $R_{IV}$ also leads to the isospin splittings of chiral phase transition boundaries and affects the susceptibilities of conserved quantities~\cite{Liu21}. It is thus expected that the couplings $R_{IV}$  may affect the equation of state of isospin asymmetric quark matter and the properties of hybrid stars. For the scalar-isovector interaction, it may result in a spinodal behavior in the EOS of the hadron-quark mixed phase and the corresponding hybrid star is unstable~\cite{Liu16}. Thus, we will mainly investigate, in this work, the role of vector and vector-isovector interactions of the quark matter in hybrid stars.

In the present study, we employ the parameters $G_S\Lambda^2 = 3.6$, $K\Lambda^5 = 8.9$, and the cutoff value in the momentum integral $\Lambda = 750$ MeV given in Refs.~\cite{Bra13,Lut92}. In the above expression, $\varepsilon_0$ is introduced to ensure $\varepsilon_{NJL}=0$ in vacuum. The pressure at zero temperature can be given as
\begin{eqnarray}
P_Q=\sum_{i=u, d, s}\mu_i\rho_i-\varepsilon_Q.
\end{eqnarray}
For leptons, we include both electrons and muons with their masses $m_e=0.511$ MeV and $m_\mu=106$ MeV, respectively. The energy density and the pressure can be given as
\begin{eqnarray}
\varepsilon_L&=&\sum_{i=e, \mu}\frac{1}{\pi^2}\int_0^{p_f^i}\sqrt{p^2+m_i^2}p^2dp,
\\
P_L&=&\sum_{i=e, \mu}\mu_i\rho_i-\varepsilon_L.
\end{eqnarray}
where $p_f^i=(3\pi^2\rho_i)^{\frac{1}{3}}$ is the lepton Fermi momentum. The total energy density and pressure including the contributions from both quarks and leptons are given by
\begin{eqnarray}
\varepsilon^Q&=&\varepsilon_Q+\varepsilon_L,
\\
P^Q&=&P_Q+P_L.
\end{eqnarray}

In the low-density hadronic phase, an improved isospin- and momentum-dependent effective nuclear interaction is used to describe the $\beta$-equilibrium and charge-neutral neutron star matter. The potential energy density from the ImMDI model is then given by~\cite{Jxu15}
\begin{eqnarray}
V_{ImMDI} &=&\frac{A_{u}\rho _n\rho _p}{\rho _{0}}
+\frac{A_{l}}{2\rho _0}(\rho_n^2+\rho_p^2)
\notag\\
&+& \frac{B}{\sigma+1} \frac{\rho^{\sigma+1}}{{\rho_0^{\sigma}}}
\times(1-x\delta^{2}) +\frac{1}{\rho_0}\sum_{\tau,\tau^{\prime}}C_{\tau,\tau^{\prime}}
\notag \\
&\times& \int\int d^3\vec{p}d^3\vec{p}^{\prime}\frac{f_{\tau}(\vec{r},\vec{p})f_{\tau^{\prime}}(\vec{r}^{\prime},\vec{p}^{\prime})}
{1+(\vec{p}-\vec{p}^{\prime })^{2}/\Lambda ^{2}}, \label{VHP}
\end{eqnarray}
where $\rho_n$ and $\rho_p$ are the neutron and proton number densities, respectively; $\rho_0= 0.16$ fm$^{-3}$ is the saturation density of nuclear matte; $\delta = (\rho_n-\rho_p)/\rho$ is the isospin asymmetry of nuclear matter with $\rho = \rho_n+\rho_p$; $f_{\tau}(\vec{r},\vec{p})$ is the nucleon phase-space distribution function from the Wigner transformation of its density matrix with $\tau= 1(-1)$ for neutrons (protons) being the isospin index. The parameter set ($A_l$, $A_u$, $B$, $C_l = C_{\tau,\tau}$ , $C_u = C_{\tau,-\tau}$, $\Lambda$, $\sigma$) can be fitted by seven empirical constraints, i.e., five isoscalar constraints of the saturation density $\rho_0$, the binding energy $E_0$, the incompressibility $K_0$, the isoscalar effective mass $m^\star_s$, and the single-particle potential $U_{0,\infty}$ at infinitely large nucleon momentum in symmetric nuclear matter, as well as two isovector constraints of the symmetry energy $E_{sym}(\rho_0)$ and the symmetry potential $U_{sym,\infty}$ at infinitely large nucleon momentum. In Ref.~\cite{Jxu15}, an optimized parameter set ($A_0$, $B$, $C_{l0}$, $C_{u0}$, $\Lambda$, $\sigma$, $x$, $y$, $z$) was introduced by using the following relations
\begin{eqnarray}
A_l(x, y)  &=& A_0 + y + x\frac{2B}{\sigma+1},
\notag\\
A_u(x, y)  &=& A_0 - y - x\frac{2B}{\sigma+1},
\notag\\
C_{\tau,\tau}(y) &=& C_{l0} - 2(y - 2z)\frac{p^2_{f0}}{\Lambda^2\text{ln}[(4p^2_{f0}+\Lambda^2)/\Lambda^2]},
\notag\\
C_{\tau,-\tau}(y)&=& C_{u0} + 2(y - 2z)\frac{p^2_{f0}}{\Lambda^2\text{ln}[(4p^2_{f0}+\Lambda^2)/\Lambda^2]},
\end{eqnarray}
where $p_{f0}$ is the nucleon Fermi momentum in symmetric nuclear matter(SNM) at saturation density. In the above relations, the parameters $x$, $y$ and $z$ are introduced to adjust the slope $L(\rho)$ of symmetry energy, the momentum dependence of the symmetry potential, and the symmetry energy $E_{sym}(\rho_0)$ at saturation density, respectively. The values of $x$, $y$, and $z$ only affect the isovector properties of nuclear matter but do not lead to the variation of the isoscalar constraints~\cite{Jxu15}. For $x = 0$, $y = 0$, and $z = 0$, we choose the following empirical values $\rho_0 = 0.16$ fm$^{-3}$, $E_0(\rho_0) = -15.9$ MeV, $K_0 = 250$ MeV, $m^{\star}_s = 0.7m$, $E_{sym}(\rho_0) = 32.5$ MeV, and $U_{0,\infty} = 75$ MeV, which lead to $A_{l0} = A_{u0} = -25.9591$ MeV, $B = 101.004 $ MeV, $C_{l0} = -60.4860$ MeV, $C_{u0} = -99.7017$ MeV, $\Lambda =2.42401p_{f0}$, and $\sigma = 1.39521$. It should be noted that the incompressibility $K_0=250$ MeV is a reasonably large value relative to the constraint $K_0 = 230\pm20 $ MeV~\cite{Shl06,Pie10}, which can stiffen the EOS of SNM so as to support more massive compact stars. And this new parametrization of the ImMDI model can be dubbed as ImMDI-ST. Recently, the discovery of GW170817 has triggered many analyses of neutron star observables, mostly the tidal deformability and radii, to constrain nuclear symmetry energy. The average value of the slope parameter of the symmetry energy $L(\rho_0)$ from the 24 new analyses of neutron star observables since GW170817 was about $L(\rho_0) = 57.7 \pm 19$ MeV at a $68\%$ confidence level~\cite{BLi21}, which is consistent with the latest report of the slope parameter $L(\rho_0)$ between 42 and 117 MeV from studying the pion spectrum ratio in heavy-ion collision in an experiment performed at RIKEN~\cite{Est21}. However, the Lead Radius Experiment (PREX-II) reported very recently new constraints on the neutron radius of $^{208}$Pb, which implies a neutron skin thickness of $R^{^{208}Pb}_{skin} = 0.283 \pm 0.071$ fm~\cite{Adh21}. From this measurement, Ref.~\cite{Ree21} constrains the slope parameter to $L(\rho_0) = 106 \pm 37$ MeV, which is much larger than many previous constraints from microscopic calculations or experimental measurements~\cite{Tsa12,Lat13,BLi21}. These new constraints can be directly compared to the inferences from gravitational wave observations of the binary compact star merger inspiral~\cite{BLi21}.

In the mean-field approximation, Eq. (10) leads to the following single-particle potential~\cite{Jxu15}
\begin{eqnarray}
U_{\tau}(\rho,\delta,\vec{p})&=& A_u\frac{\rho_{-\tau}}{\rho_0}+A_l\frac{\rho_{\tau}}{\rho_0}
\notag\\
&+& B\frac{\rho}{\rho_0}^{\sigma}(1-x\delta^2)- 4x\tau\frac{B}{\sigma+1}\frac{\rho^{\sigma-1}}{\rho_0^{\sigma}}\delta\rho_{-\tau}
\notag\\
&+&\frac{2C_l}{\rho_0}\int d^3\vec{p}^{\prime}\frac{f_{\tau}(\vec{r},\vec{p})}{1+(\vec{p}-\vec{p}^{\prime })^{2}/\Lambda ^{2}}
\notag\\
&+&\frac{2C_u}{\rho_0}\int d^3\vec{p}^{\prime}\frac{f_{-\tau}(\vec{r},\vec{p})}{1+(\vec{p}-\vec{p}^{\prime })^{2}/\Lambda ^{2}}.
\end{eqnarray}

The chemical potential of neutrons and protons can be calculated from
\begin{equation}
\mu_{\tau }=\sqrt{m^2+p_f^{\tau2}}+ U_\tau(p_f^{\tau}),
\end{equation}
with the nucleon mass $m$ and the Fermi momentum $p_f^{\tau}=(3\pi^2\rho_\tau)^{1/3}$. The total energy density and pressure of the hadron phase can be written as
\begin{eqnarray}
\varepsilon^H&=&\varepsilon_H+\varepsilon_L,
\\
P^H&=&P_H+P_L,
\end{eqnarray}
where $\varepsilon_H$ and $P_H$, respectively, are energy density and pressure of baryons. The detailed form can be written as
\begin{eqnarray}
\varepsilon_H&=&V_{HP}+V_{HK}+V_{HM},
\notag \\
P_H&=&\sum_{\tau}\mu_{\tau}\rho_{\tau}-\varepsilon_H,
\end{eqnarray}
where $V_{HP}$ is the potential energy density of baryons calculated from $V_{ImMDI}$,
$V_{HK}$ and $V_{HM}$ are, respectively, the kinetic energy and mass contributions given by
\begin{eqnarray}
V_{HK}&=&\sum_{\tau}\frac{p_f^{\tau5}}{10\pi^2m_{\tau}},
\notag \\
V_{HM}&=&\sum_{\tau}\rho_{\tau}m_{\tau}.
\end{eqnarray}

At moderate densities of hybrid stars, the hadron-quark phase transition, which leads to a mixed phase of hadronic and quark matter, can be described by the Gibbs conditions~\cite{GLe92, GLe01}
\begin{eqnarray}
T^H&=&T^Q, \qquad \qquad P^H=P^Q,
\notag\\
\mu_B&=&\mu_B^H=\mu_B^Q,  \quad \mu_c=\mu_c^H=\mu_c^Q.
\end{eqnarray}
Adding baryon number conservation, and charge neutrality conditions, the dense matter enters the mixed phase, in which the hadron phase and the quark phase need to satisfy following equilibrium conditions:
\begin{eqnarray}
\mu_i&=&\mu_Bb_i-\mu_cq_i,  \quad P^H=P^Q,
\notag\\
\rho_B&=&(1-Y)(\rho_n+\rho_p)+\frac{Y}{3}(\rho_u+\rho_d+\rho_s),
\notag\\
0&=&(1-Y)\rho_p+\frac{Y}{3}(2\rho_u-\rho_d-\rho_s)-\rho_e-\rho_\mu,
\end{eqnarray}
where $Y$ is the baryon number fraction of the quark phase. The total energy density and pressure
of the mixed phase are calculated according to
\begin{eqnarray}
\varepsilon^M&=&(1-Y)\varepsilon_H+Y\varepsilon_Q+\varepsilon_L,
\\
P^M&=&(1-Y)P_H+YP_Q+P_L.
\end{eqnarray}

Besides, in our calculations, the crust of hybrid stars is considered to be divided into two parts: the inner and the outer crust as in the previous treatment~\cite{Jxu09C, Jxu09J}. In the inner crust, a parametrized EOS of $P = a + b\varepsilon^{4/3}$ is used and the outer crust usually consists of heavy nuclei and an electron gas, where we use the EOS in Ref.~\cite{Bay71}.

The whole EOS from low densities to high densities is used to study the mass-radius relation of hybrid stars through the Tolman-Oppenheimer-Volkoff (TOV) equations and the analytical expression of TOV equations can be written as
\begin{eqnarray}
\frac{dP(r)}{dr}&=&-\frac{M(r)[\varepsilon(r)+P(r)]}{r^2}[1+\frac{4 \pi P(r)r^3}{M(r)}]
\notag\\
&\times&[1-\frac{2M(r)}{r}]^{-1},
\end{eqnarray}
where $\varepsilon(r)$ is the energy density and $P(r)$ is the pressure obtained from the equation of state. $M(r)$ is the gravitational mass inside the radius $r$ of the compact star which can be obtained from the integral of the following equation
\begin{eqnarray}
\frac{dM(r)}{dr}&=&4 \pi r^2 \varepsilon(r).
\end{eqnarray}

Coalescing binary compact stars is one of the most promising sources of gravitational waves. One of the most important
features of binary mergers is the tidal deformation, which is considered as another probe to the EOS of dense matter~\cite{Hin08,Rea09}. The tidal deformability $\Lambda$ of compact stars during their merger is related to the Love number $k_2$ through the relation $k_2 =3/2\Lambda\beta^5$~\cite{Hin08,Pos10}, which can be given by
\begin{eqnarray}
k_2 &=&\frac{8}{5}(1-2\beta)^2[2-y_R+2\beta(y_R-1)]
\notag\\
&\times& \{2\beta[6-3y_R+3\beta(5y_R-8)]
\notag\\
&+& 4\beta^3[13-11y_R+\beta(3y_R-2)+2\beta^2(1+y_R)]
\notag\\
&+&3(1-2\beta)^2[2-y_R+2\beta(y_R-1)]\text{ln}(1-2\beta)\}^{-1},
\end{eqnarray}
where $\beta \equiv M/R $ is the compactness of the compact star, and $y_R \equiv y(R)$ is the solution at the compact star surface to the first
order differential equation
\begin{eqnarray}
r\frac{dy(r)}{dr}+y(r)^2+y(r)F(r)+r^2Q(r)=0,
\end{eqnarray}
with
\begin{eqnarray}
F(r) &=& \frac{r-4\pi r^3[\varepsilon(r)-P(r)]}{r-2M(r)},
\notag\\
Q(r)&=&\frac{4\pi r[5\varepsilon(r)+9P(r)+\frac{\varepsilon(r)+P(r)}{\partial P(r)/\partial \varepsilon(r)}-\frac{6}{4\pi r^2}]}{r-2M(r)}
\notag\\
&-&4[\frac{M(r)+4\pi r^3P(r)}{r^2(1-2M(r)/r)}]^2.
\end{eqnarray}
For a given central density $\rho_c$ and using the boundary conditions in terms of $y(0) = 2$, $P(0)=P_c$, $M(0)=0$ and $\epsilon(0)=0$, the mass $M$, radius $R$, and the tidal deformability $\Lambda$ can be obtained once an EOS is supplied.

\section{Results and Discussions}
\label{RAD}
Before discussing the symmetry energy of hybrid star matter, we first review the symmetry energy of nuclear matter. It is well known that the binding energy of asymmetric nucleonic matter (ANM) of isospin asymmetry $\delta$ and density $\rho$ can be written as
\begin{eqnarray}
E(\rho,\delta)=E_0(\rho)+E_{sym}(\rho)\delta^2+\mathcal{O}(\delta^4),
\end{eqnarray}
where $E_0(\rho)$ and $E_{sym}(\rho)$ are the energy per nucleon in symmetric nucleonic matter (SNM) and nuclear symmetry energy, respectively. In Eq. (27), there are no odd-order terms due to the exchange symmetry between protons and neutrons in nuclear matter. The higher-order terms are generally negligibly small~\cite{Lag81,Bom91}, and the symmetry energy $E_{sym}(\rho)$ is expressed by definition as
\begin{eqnarray}
E_{sym}(\rho)=\frac{1}{2!}\frac{\partial^2E(\rho,\delta)}{\partial\delta^2}|_{\delta=0}.
\end{eqnarray}

\begin{figure}[tbh]
\includegraphics[scale=0.28]{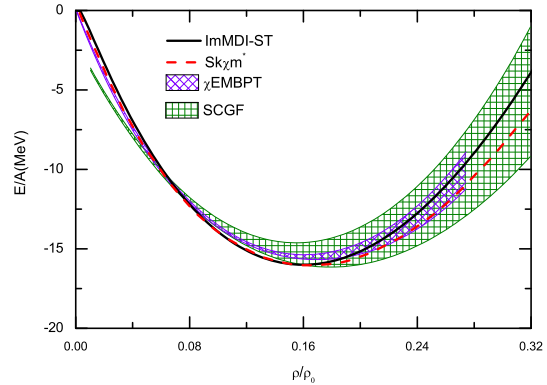}
\caption{(color online) The energy per nucleon as a function of the reduced nucleon density for symmetric nuclear matter at zero temperature from ImMDI-ST and Sk$\chi$m$^*$ compared with results from the SCGF approach and the $\chi$EMBPT approach.} \label{fig1}
\end{figure}

As stated in the introduction, we describe nuclear matter in the ImMDI model, which is constructed by fitting empirical nucleon optical potential and cold nuclear matter properties at saturation density. Again, the ImMDI-ST model is also fitted to the empirical properties of SNM, which is approximately reproduced by the self-consistent Green¡¯s function (SCGF) approach~\cite{Car14,Car18} or chiral effective many-body pertubation theory ($\chi$EMBPT)~\cite{Wel15,Wel16}. As shown in Fig.~\ref{fig1}, the density dependence of the energy per nucleon for SNM at zero temperature from the ImMDI-ST model is compared with the results from the SCGF approach and the $\chi$EMBPT approach. The result of red dash line is from the Skyrme-Hartree-Fock (SHF) model~\cite{Che10,Dut12} using the Sk$\chi$m$^*$ interaction. The two effective interactions ImMDI-ST and Sk$\chi$m$^*$ are both based on Hartree-Fock calculations and constructed from fitting the properties of cold nuclear matter. It is seen that the EOSs from ImMDI-ST and Sk$\chi$m$^*$ are almost identical for zero temperature SNM at low densities, while the two lines start to deviate around saturation density, which is due to the enhancement of the incompressibility ($K_0=250$ MeV) in the ImMDI-ST interaction. Except for small deviations at very low densities, the EOSs from these two effective interactions are within the SCGF uncertainty band, which is caused by the different momentum cutoffs and the phenomenology in three-body forces~\cite{Car14,Car18}. Fig.~\ref{fig1} also displays the results from $\chi$EMBPT calculations using n3lo414 chiral forces, which are taken from Ref.~\cite{Wel15}. Compared with the SCGF approach, the EOS of SNM from the $\chi$EMBPT is seen to be better reproduced by ImMDI-ST and Sk$\chi$m$^*$.

\begin{figure*}[tbh]
\includegraphics[scale=0.45]{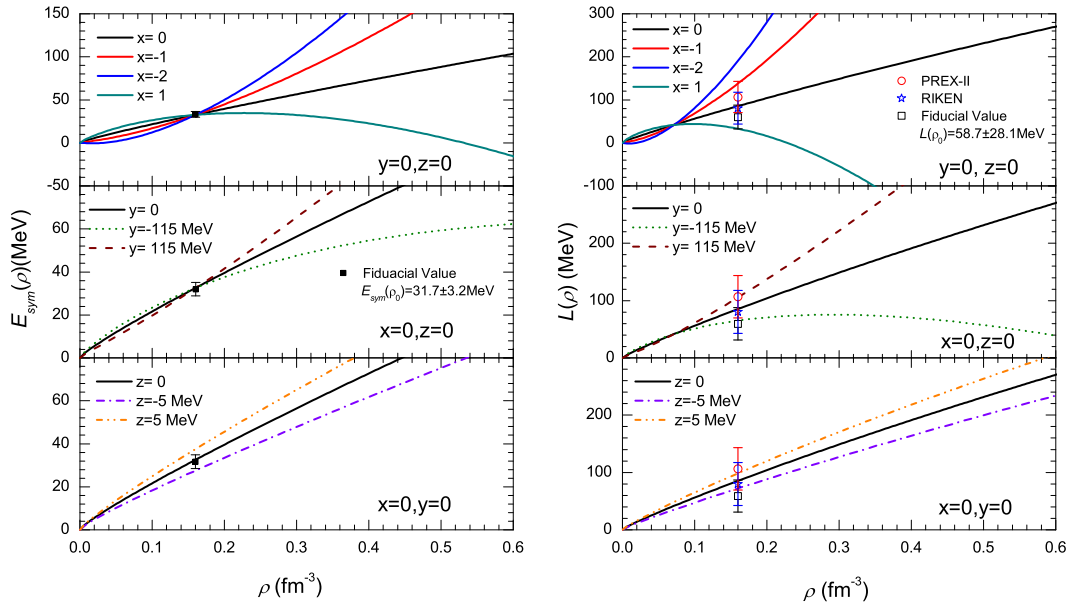}
\centering
\caption{(color online) The nuclear symmetry energy $E_{sym}(\rho)$ (left) and its slope $L(\rho)$ (right) as functions of the nucleon density $\rho$ from the ImMDI-ST interaction with varying parameters $x$, $y$ and $z$. The constraints on $E_{sym}(\rho_0)$ and $L(\rho_0)$ from (1) the fiducial value of 53 analyses about nuclear experiments and astrophysical observations in 2016~\cite{Wat16}, (2) studying the pion spectrum ratio in heavy-ion collision in an experiment performed at RIKEN~\cite{Est21}, (3) the PREX-II experiment based on the relativistic mean-field (RMF) model calculations~\cite{Adh21} are also shown for comparison.}\label{fig2}
\centering
\label{EOS}
\end{figure*}

For asymmetric nucleonic matter (ANM), as shown in Eq. (27), symmetry energy is an important part of the EOS of nuclear matter. In the ImMDI-ST interaction, one can adjust flexibly three parameters ($x$, $y$, and $z$) to change the isospin properties of nuclear matter. In fact, the values of $x$, $y$, and $z$ only affect the isovector properties of nuclear matter without leading to the variation of properties of SNM. This phenomena is illustrated in Fig.~\ref{fig2} which displays the density dependence of nuclear symmetry energy and its slope using the ImMDI-ST interaction by varying the parameter $x$, $y$, and $z$. Different values of $x$ can lead to widely different trends for the symmetry energy $E_{sym}(\rho)$ and the slope $L(\rho)$ while the magnitude of the symmetry energy remains unchanged at saturation density. Qualitatively, both $E_{sym}(\rho)$ and $L(\rho)$ decrease with the increment of the parameter $x$ at supra-saturation density. On the other hand, it is also seen that the density dependence of the symmetry energy and its slope change with the parameter $y$ as well, which is due to that the parameter $y$ can modify the momentum dependence of the symmetry potential $U_{sym}(\rho, p)$~\cite{Cxu10,Che12}. In addition, the value of the symmetry energy at saturation density can be adjusted by parameter $z$, which also affects the behavior of the symmetry energy and its slope at nonsaturation density.  In Fig.~\ref{fig2}, the fiducial value of $E_{sym}(\rho_0) = 31.7 \pm 3.2$ MeV and $L(\rho_0) = 58.7 \pm 28.1$ MeV from the 2016 survey of 53 analyses about nuclear experiments and astrophysical observations~\cite{Wat16}, the slope parameter $L(\rho_0)$ between 42 and $117$ MeV from studying the pion spectrum ratio in heavy-ion collision in an experiment performed at RIKEN~\cite{Est21}, as well as the value of $L(\rho_0) = 106 \pm 37$ MeV~\cite{Adh21} based on the PREX-II experiment are also shown for comparison. Except the value of the slope with $x=-2$, the rest values of the nuclear symmetry energy and its slope at saturation density are approaching these empirical constraints.

\begin{figure*}[tbh]
\includegraphics[scale=0.45]{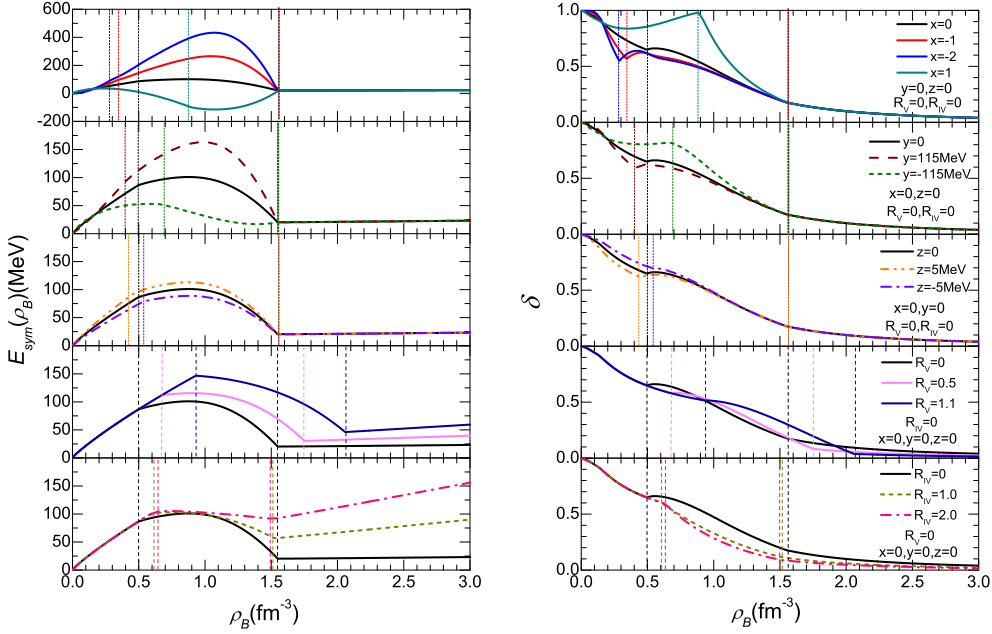}
\caption{(color online) The symmetry energy $E_{sym}(\rho_B)$ (left) and the corresponding isospin asymmetry $\delta$ (right) in hybrid stars at $\beta$ equilibrium as functions of the baryon density by varying the parameters $x$, $y$ and $z$ from ImMDI-ST interactions as well as the couplings $R_V$ and $R_{IV}$ from NJL model. The two dashed lines with the same color indicate the range of the mixed phase (the hadron-quark phase transition).} \label{fig3}
\end{figure*}

Although researches based on terrestrial nuclear experiments and astrophysical observations have already led to many significant constraints on the EOS of symmetric nuclear matter(SNM) and the symmetry energy $E_{sym}(\rho)$ around the saturation density, huge uncertainties remain at higher densities. The radii and tidal deformations of compact stars are considered to be determined by the pressure around the density $2\rho_0$, which provides the possibility of constraining $E_{sym}(\rho)$ at higher densities. Moreover, many theories predict that at densities higher than about $(2 \sim 5)\rho_0$~\cite{Ann20,Sha13,Ors13,Wu19,Mju21}, a hadron-quark phase transition will occur. Since $E_{sym}(\rho)$ will lose its physical meaning once the hadron-quark phase transition happens, one thus has to redefine the symmetry energy and the isospin asymmetry manifesting isospin properties of quark matter. In the present study, we investigate the properties of symmetry energy at higher densities, especially the density $2\rho_0$, in hybrid star matter with the hadron-quark phase transition.
The symmetry energy in the hybrid star matter can be defined as
\begin{eqnarray}
E_{sym}(\rho_B,\rho_s)=\frac{1}{2!}\frac{\partial^2E(\rho_B,\delta,\rho_s)}{\partial\delta^2}|_{\delta=0},
\end{eqnarray}
where $E(\rho_B,\delta,\rho_s)$ is the energy per baryon number for isospin asymmetric matter, and one can obtain a similar definition of the symmetry energy for quark matter in Refs.~\cite{BLi08,Xli15}.  In the above, $\rho_s$ stands for the strange quark number density and $\rho_B$ means the baryon number density defined by $\rho_B = (1-Y)(\rho_n-\rho_p)+\frac{Y}{3}(\rho_u+\rho_d+\rho_s)$, where $Y$ is the baryon number fraction of the quark phase. Similarly, the isospin asymmetry in the hybrid star matter can be defined as
\begin{eqnarray}
\delta =\frac{(1-Y)(\rho_n-\rho_p)+Y(\rho_d-\rho_u)}{\rho_B}.
\end{eqnarray}

Isospin properties of quark matter in hybrid star matter can be obtained from 3-flavor NJL model with the vector and vector-isovector interactions. As discussed in the above, the position of the critical point for the chiral phase transition is sensitive to the vector coupling constants $R_V$, which also helps to explain the elliptic flow splittings between protons and antiprotons in RHIC-BES experiments. The vector-isovector coupling constant $R_{IV}$ leads to different potentials of $u$ and $d$ quarks in isospin asymmetric quark matter and the isospin splittings of chiral phase transition boundaries~\cite{Liu16,Fra03,Zha14}. They are thus expected to affect the symmetry energy and the EOS of the hybrid star matter. In Fig.~\ref{fig3}, we show the symmetry energy $E_{sym}(\rho_B)$ (left panel) of the hybrid star matter at $\beta$ equilibrium as functions of the baryon density by varying the parameters $x$, $y$, and $z$ from the ImMDI-ST model as well as the coupling constants $R_V$ and $R_{IV}$ from the NJL model. The two dash lines with the same color indicate the range of the mixed phase (the hadron-quark phase transition). It can be seen that the parameters $x$, $y$, and $z$ affect the hadronic phase and the mixed phase, but the coupling constants $R_V$ and $R_{IV}$ affect the mixed phase and the quark phase. All of the parameters and coupling constants of the interactions from the ImMDI-ST and NJL model can lead to widely different trends for the symmetry energy in the mixed phase and positions of the onset of the hadron-quark phase transition. In detail, decreasing the value of $x$ effectively stiffens the symmetry energy in the mixed phase while the results with the parameters $y$ and $z$ demonstrate oppositely, which is similar to the results from the nuclear symmetry energy. Furthermore, $E_{sym}(\rho_B)$ in the mixed phase and the quark phase increases with the increment of both the coupling constants $R_V$ and $R_{IV}$. On the other hand, it is also seen that the stiffened symmetry energy for the ImMDI-ST interaction can increase the baryon density of the hadron-quark phase transition point, while the stiffened symmetry energy for the interaction within NJL model can decrease the onset to the lower baryon densities.  In addition, with various parameters and coupling constants, the hadron-quark phase transition appears around $2\rho_0 \thicksim 6\rho_0$, where the radii, masses and tidal deformations of compact stars are considered to are sensitive to the density dependence of symmetry energy. The right panel of the Fig.~\ref{fig3} displays the corresponding isospin asymmetry $\delta$ in hybrid stars at $\beta$-equilibrium with varying the parameters and the coupling constants. It can be seen that a non-smooth local extremum appears at the onset of the hadron-quark phase transition in all cases, which is due to the occurrence of the first-order phase transition and the appearance of $s$ quark that change the $\beta$-equilibrium conditions at zero temperature. As the $E_{sym}(\rho_B)$ varies broadly with the different parameters and coupling constants in the mixed phase, one also can see that the value of the corresponding $\delta$ for neutron-rich matter is obtained with the soft $E_{sym}(\rho_B)$ and that for neutron-poor matter is obtained with the stiff $E_{sym}(\rho_B)$.

\begin{figure}[tbh]
\includegraphics[scale=0.28]{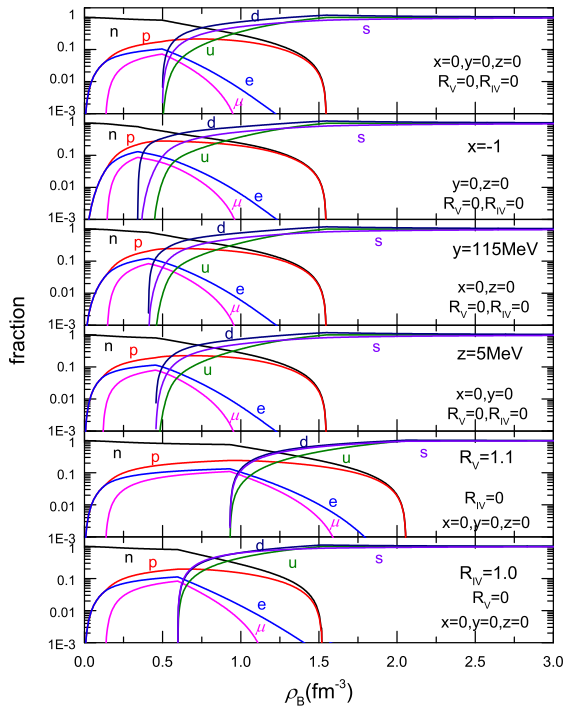}
\caption{(color online) Relative particle fractions of hadrons, leptons and quarks as functions of baryon density in hybrid stars for the different parameters $x$, $y$, and $z$ from ImMDI-ST interactions as well as coupling constants $R_V$ and $R_{IV}$ from NJL model.} \label{fig4}
\end{figure}

To better understand the properties of the hybrid star matter, we show in Fig.~\ref{fig4} the relative particle fractions in hybrid stars with the different parameters and coupling constants. It can be observed in all cases that the fractions of the neutrons and leptons both decrease while that of the protons increases when the hadron-quark phase transition occurs, so $d$ and $s$ quarks occupy a larger fraction than $u$ quarks in the mixed phase in order to maintain electrical neutrality. With the decrease of proton fraction and the disappearance of electrons at high densities, the isospin asymmetry of the $d$ and $u$ quark gradually decreases to zero. For comparison, we choose the following values $x=-1$, $y=115$ MeV, $z=5$ MeV, $R_V=1.1$, and $R_{IV}=1.0$, which all cause the symmetry energy to be stiffer. It is also observed that quarks generally appear at lower densities for the ImMDI-ST parameters $x=-1$, $y=115$ MeV, and $z=5$ MeV, while it is the opposite for the quark coupling strengths $R_V=1.1$ and $R_{IV}=1.0$. Then a fatal problem for the hybrid star matter  will occur here, that is, if the EOS of nuclear matter is stiffer, the softer quark matter appears earlier, while if the EOS of quark matter is stiffer, the nuclear matter will disappear later. Furthermore, the mixed phase always seems to favor to be in a softer equation of state,and that is the reason why it is difficult to reproduce a more massive hybrid star with the $\beta$-equilibrium condition.

\begin{figure}[tbh]
\includegraphics[scale=0.28]{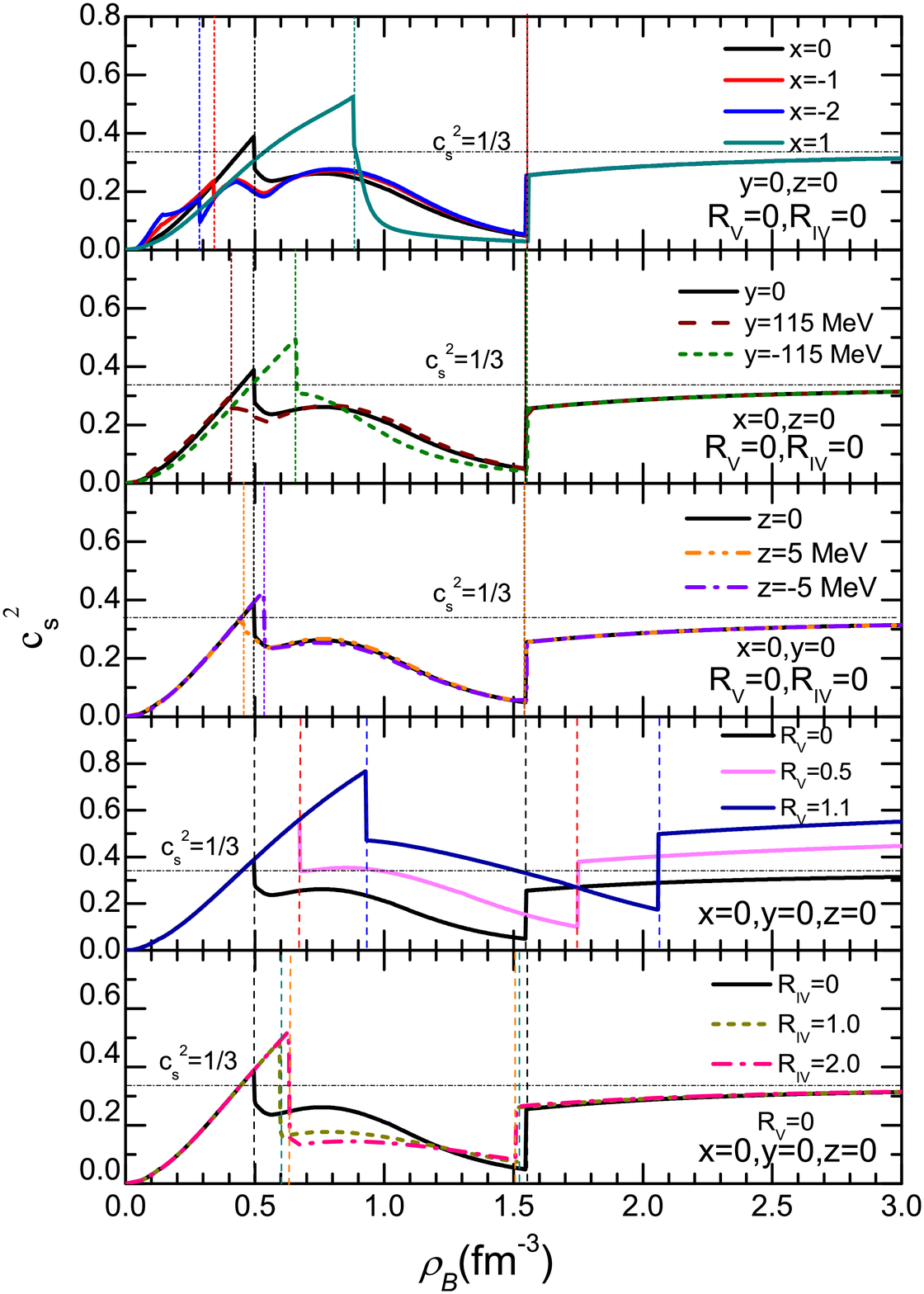}
\caption{(color online) The square of the sound velocity as a function of baryon density in hybrid stars for the different parameters $x$, $y$, and $z$ from ImMDI-ST interactions as well as coupling constants $R_V$ and $R_{IV}$ from the NJL model.} \label{fig5}
\end{figure}

The sound velocity $c_s$, which can be calculated from $c^2_s=\partial P/\partial \epsilon$, is another property of the hybrid star matter, and it can be used to check if the underlying EOS satisfies the causality condition. As shown in Fig.~\ref{fig5}, the sound velocity is sensitive to the nuclear matter interactions at low densities and to the quark matter interactions at high densities. As expected, a stiffer symmetry energy leads to a larger value of the sound velocity in the hadron and quark phase. However, a step change of the sound velocity occurs in the mixing phase where the quarks appear and thus soften the EOS as a result of more degrees of freedom, and it is restored with the decrease of nucleon and lepton degrees of freedom in the high density quark phase. Also shown in Fig.~\ref{fig5} is the sound velocity $c^2_s=1/3$ in the conformal limit corresponding to free massless fermions, and it is seen that our results with a strong repulsive vector interaction for quark matter are larger than this limit at higher densities, indicating that the corresponding EOS is stiffer than that of massless fermions. For the vector-isovector interaction case, the result is quite different, which is due to the contribution from the $G_{IV}(\rho_u-\rho_d)^2$ term in Eq. (4) (this term is obviously sensitive to the isospin asymmetry at higher densities). Meanwhile, we note that for all cases considered here the causality condition is safely satisfied.

\begin{figure*}[tbh]
\includegraphics[scale=0.3]{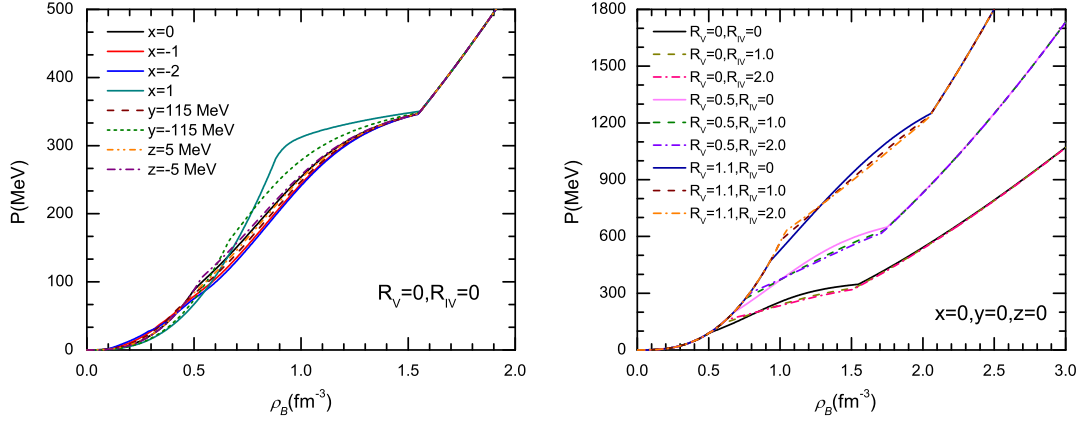}
\caption{(color online) EOSs of hybrid stars based on the ImMDI-ST interactions for nuclear matter with different parameters $x$, $y$, and $z$ (left) as well as the NJL model for quark matter with different coupling constants $R_V$ and $R_{IV}$ (right).} \label{fig6}
\end{figure*}

We present in Fig.~\ref{fig6} the EOS for hybrid stars with the hadron-quark phase transition in their inner core. Again, the results from the ImMDI-ST interactions for nuclear matter with different parameters $x$, $y$, and $z$ (left panel) as well as those from the NJL model for quark matter with different coupling constants $R_V$ and $R_{IV}$ (right panel) are shown. It can be seen in left panel that a stiff symmetry energy for nuclear matter based on ImMDI-ST model, i.e. $x=-1$, $y=115$ MeV, and $z=5$ MeV, leads to a stiffer EOS in hadron phase at low densities and a slower increase with increasing baryon density in the mix phase, so that all curves represented different ImMDI-ST interactions can enter the pure quark phase at same density. These are understandable since the ImMDI-ST interactions tend to increase the symmetry energy of nuclear matter and have no effects on pure quark matter. Moreover, as shown in the right panel of Fig.~\ref{fig6}, the EOS is more sensitive to the strength of the vector interaction through the $G_V(\rho_u^2+\rho_d^2+\rho_s^2)$ term in Eq.(4). With increasing vector strength $R_V$ for the strange quark matter the EOS of hybrid star becomes stiffer, which is consistent with that observed in Ref.~\cite{Han01}, and the onset of the transition is moving to higher densities. The vector-isovector interaction characterized by the coupling constant $R_{IV}$ slightly stiffens the EOS at low densities in the mixed phase, since its contribution is determined by the $G_{IV}(\rho_u-\rho_d)^2$ term in Eq.(4). Similar to the effect of the ImMDI-ST interactions, the effect of vector-isovector interaction also decreases gradually at high densities in the mixed phase, which is due to the decrease of isospin asymmetry as shown in Fig.~\ref{fig2}.

\begin{figure*}[tbh]
\includegraphics[scale=0.3]{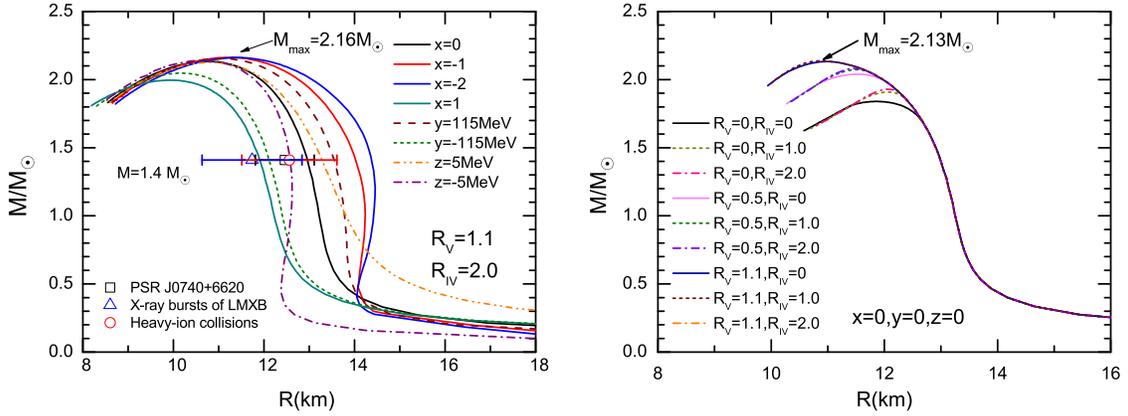}
\caption{(color online) Mass-radius relation of hybrid stars based on the ImMDI-ST interactions for nuclear matter with different parameters $x$, $y$ and $z$ (left) as well as the NJL model for quark matter with different coupling constants $R_V$ and $R_{IV}$ (right). The constraints of $10.62 \leq R_{1.4} \leq 12.83$ km from the x-ray bursts of accreting compact stars in low-mass x-ray binary (LMXB) systems~\cite{Lat14}, $12.45\pm0.65$ km of PSR J0740+6620 for a 1.4 $M_{\odot}$ compact star based on Neutron Star Interior Composition Explorer (NICER) and X-ray Multi-Mirror (XMM-Newton)~\cite{Mil21} as well as the prediction of $11.5 \leq R_{1.4} \leq 13.6$ km from heavy-ion collisions~\cite{BLi06} are also shown for comparison.} \label{fig7}
\end{figure*}

Essentially all available EOSs can be used to predict the mass-radius correlation of compact stars. Many of the earlier studies have focused on exploring the effects of the properties of SNM and symmetry energy near the saturation density. Effects of varying the $E_{sym}$ in hybrid star matter will be studied extensively in this work. We show in Fig.~\ref{fig7} the mass-radius relation of hybrid stars based on the ImMDI-ST interactions for nuclear matter and the NJL model for quark matter. The results shown in the left panel indicate that the observed maximum mass of hybrid stars change slightly with the different ImMDI-ST interactions. Except for the cases $x=1$ and $y=-115$ MeV, the maximum mass of all other hybrid stars is very close to the detection result of the MSR J0740+6620($2.14^{+0.20}_{-0.18} M_{\odot}$ with 95.4\% credibility) with the quark coupling constants fixed on $R_V=1.1$ and $R_{IV}=2.0$, which is slightly larger than the measurement of PSR J0740+6620 with the gravitational mass $2.08 \pm 0.07M_{\odot}$. This is due to that the maximum mass of hybrid stars constrains mostly the EOS of symmetric matter at $2\sim5$ times saturation density. As mentioned earlier, the different parameters $x$, $y$, and $z$ in the ImMDI-ST interactions will not lead to the variation of properties of SNM, and thus have no significant effect on the maximum mass of hybrid stars. However, it should be noted that the hadron-quark transition in most cases (except for $x=1$ and $y=-115$ MeV) occurs at $2\sim5$ times saturation density, thus the properties of symmetric quark matter will affect the maximum mass of hybrid stars. This is illustrated in the right panel of Fig.~\ref{fig7} which displays mass-radius relation of hybrid stars with different quark coupling constants $R_V$ and $R_{IV}$. Compared to the vector-isovector interaction, the vector interaction of quark matter plays an more important role in mass-radius relations of hybrid stars. With increasing vector interaction coupling $R_V$ in the quark matter hybrid stars have larger maximum mass. This can be understandable since the vector interaction affects the EOS of not only asymmetric quark matter but also symmetric quark matter through the $G_V(\rho_u^2+\rho_d^2+\rho_s^2)$ term in Eq.(4).

On the other hand, it is clearly seen from Fig.~\ref{fig7} that the radii are more sensitive to the parameters of ImMDI-ST interactions, which is due to that the radii of hybrid stars constrain mostly the density dependence of symmetry energy. This is also explained in Ref.~\cite{BLi06M} by studying the relative contributions from the SNM EOS and symmetry energy to the total pressure in neutron stars at $\beta$-equilibrium. The results from Ref.~\cite{BLi06M} indicate that in the density region around $\rho_0 \sim 2.5 \rho_0$, the isospin dependent
pressure $P_{asy}$ dominates over the $P_0$ from SNM, while the total pressure is dominated by the $P_0$ from SNM at higher densities. The radii of compact stars are known to be determined by the pressure at densities around $\rho_0\sim 2\rho_0$~\cite{Lat00,Lat01}. Thus, the radii of hybrid stars constrain mostly the density dependence of symmetry energy while observed maximum mass of hybrid stars constrain mostly the EOS of symmetric matter. In recent reports, the x-ray bursts from accreting compact stars in low-mass x-ray binary (LMXB) systems provide potential possibilities to constrain the mass and radius simultaneously. As summarized, several constraints on the radii of neutron stars have been put forward in recent years: $10.4 \leq R_{1.4} \leq 12.9$ km~\cite{Ste13}, $10.62 \leq R_{1.4} \leq 12.83$ km~\cite{Lat14}, $10.1 \leq R_{1.4} \leq 11.1$ km~\cite{Oze16}, $10.6 \leq R_{1.4} \leq 14.2$ km~\cite{Sha18}, $10 \leq R_{1.4} \leq 14.4$ km~\cite{Ste18}. For comparison, the constraint of $10.62 \leq R_{1.4} \leq 12.83$ km~\cite{Lat14} is shown in Fig.~\ref{fig7}. Besides, the radius measurement $12.45 \pm 0.65$ km of PSR J0740+6620 for a $1.4 M_\odot$ compact star based on Neutron Star Interior Composition Explorer (NICER) and X-ray Multi-Mirror (XMM-Newton)~\cite{Mil21} as well as the prediction of $11.5 \leq  R_{1.4} \leq  13.6$ km~\cite{BLi06} from heavy-ion collisions are also shown in the figure. Meanwhile, we also can see in the right panel of Fig.~\ref{fig7} that isospin properties of quark matter have no effect on the radii of $1.4 M_\odot$ hybrid stars as a result of no quark matter at densities around $\rho_0\sim 2\rho_0$.

\begin{figure*}[tbh]
\includegraphics[scale=0.3]{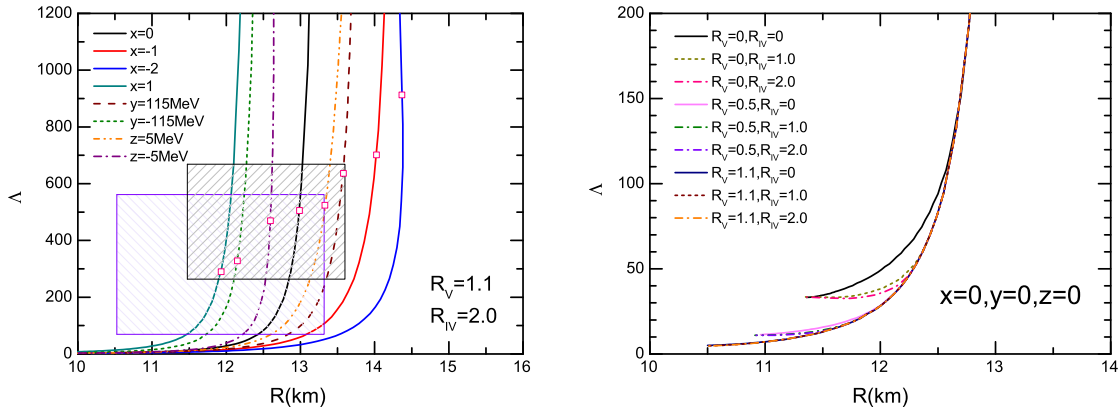}
\caption{(color online) Relation between the dimensionless tidal deformability and the radius of hybrid star based on the ImMDI-ST interactions for nuclear matter with different parameters $x$, $y$ and $z$ (left) as well as the NJL model for quark matter with different coupling constants $R_V$ and $R_{IV}$ (right). The squared violet region of $70 \leq \Lambda_{1.4} \leq 580$ and $10.5 \leq R_{1.4} \leq 13.3$ km corresponds to the constraints reported by the LIGO and Virgo Collaborations~\cite{Abb18}, while the squared black region of $292 \leq \Lambda_{1.4} \leq 680$ and $11.5 \leq R_{1.4} \leq 13.6$ km corresponds to the constraints from heavy-ion collisions~\cite{BLi06}. And the small pink squares indicate the results for hybrid stars with $M = 1.4M_{\odot}$.} \label{fig8}
\end{figure*}

After the GW170817 event, much efforts have been devoted to constraining the EOS or related model parameters by comparing various calculations with the range of tidal deformability $70 \leq \Lambda \leq 580$ from the improved analyses reported by LIGO and Virgo Collaborations. A number of these studies have examined the effects of symmetry energy~\cite{Kra19,Car19,Zha19}. Some of them have extracted constraints on the slope parameter $L(\rho_0)$, i.e. $L(\rho_0) = 57.7 \pm 19$ MeV~\cite{BLi21}. The measurements of the tidal deformability of neutron stars constrain not only the EOS of dense neutron-rich nuclear matter but also the fundamental strong interactions of quark matter. Shown in Fig.~\ref{fig8} is the dimensionless tidal deformability as functions of radius calculated using the different interactions from ImMDI-ST and NJL model. In the left panel the small pink squares indicate the results for hybrid stars with $M = 1.4M_{\odot}$. For a given mass $M = 1.4M_{\odot}$, the deformability $\Lambda_{1.4}$ increases with increasing radius of hybrid star, as expected. The radii of hybrid stars are known to be determined by the symmetry energy, thus the deformability of hybrid stars can be used to constrain the density dependence of symmetry energy. For comparison, we display some constraints in the left panel of Fig.~\ref{fig8} where the squared violet region of $70 \leq \Lambda_{1.4} \leq 580$ and $10.5 \leq R_{1.4} \leq 13.3$ km corresponds to the constraints reported by the LIGO and Virgo Collaborations~\cite{Abb18}, while the squared black region of $292 \leq \Lambda_{1.4} \leq 680$ and $11.5 \leq R_{1.4} \leq 13.6$ km corresponds to the constraints from heavy-ion collisions~\cite{BLi06}. Except for the cases $x=-1$, $x=-2$ and $y=115$ MeV, it can be seen that the $\Lambda_{1.4}$ with various parameters of ImMDI-ST interactions are approaching the overlapping part of the two constraints. The relation between the tidal deformability and the mass of hybrid stars using NJL model by varying the coupling constants $R_V$ and $R_{IV}$ is shown in the right panel. One can be seen that the vector and vector-isovector interactions have slightly effects on the minimum deformability which is related to the difference in maximum mass of hybrid stars. Similarly, these interactions have no effect on the $\Lambda_{1.4}$ of hybrid stars.

\begin{figure*}[tbh]
\includegraphics[scale=0.3]{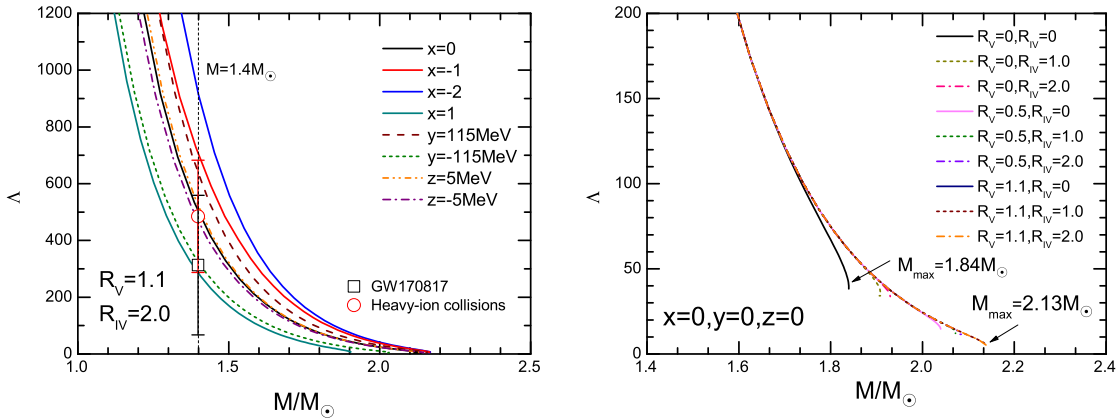}
\caption{(color online)  Relation between the dimensionless tidal deformability and the mass of hybrid star based on the ImMDI-ST interactions for nuclear matter with different parameters $x$, $y$ and $z$ (left) as well as the NJL model for quark matter with different coupling constants $R_V$ and $R_{IV}$ (right). The error bar at $1.4M_{\odot}$ corresponds to the constraints on the tidal deformability $70 \leq \Lambda \leq 580$ based on the improved analysis of GW170817 by LIGO and Virgo Collaborations as well as the prediction of $292 \leq \Lambda \leq 680$ from heavy-ion collisions.} \label{fig9}
\end{figure*}

Meanwhile, we also show in Fig.~\ref{fig9} the relation between the dimensionless tidal deformability and the mass of hybrid star. We see that $\Lambda$ decreases rapidly as the mass of the neutron star increases. This is due to the factor that given the smaller range of allowed radii for larger massive stars, the spread in the tidal deformability is also naturally much tighter than for lower-mass neutron stars. The error bar at $1.4M_{\odot}$ corresponds to the constraints on the tidal deformability $70 \leq \Lambda \leq 580$ based on the improved analysis of GW170817 by LIGO and Virgo Collaborations as well as the prediction of $292 \leq \Lambda \leq 680$ from heavy-ion collisions. It is seen that the main contribution to $\Lambda_{1.4}$ is from the hadron phase and the upper limit of $\Lambda_{1.4} = 580$ is an important constraint on the parameters $x$, $y$ and $z$. It is also seen in Fig.~\ref{fig9} that the tidal deformability is not sensitive to the vector and vector-isovector interactions. Quark matter interactions have effect on the EOS and symmetry energy of the hadron-quark mixed phase transition and thus lead to the difference of the maximum mass of hybrid stars. For the parameters $x=y=z=0$, the maximum mass of hybrid stars can reach about $2.13 M_\odot$ with $R_V=1.1$ and $R_{IV}=2.0$. Furthermore, if quark matter would make more contribution to the tidal deformability, the equation of state of nuclear matter need to be stiffer so that the quark matter appears in the low-density region.

\section{Summary and Outlook}
\label{viscosity}
The density dependence of symmetry energy is an important part of the equation of state of isospin symmetry matter. However, the huge uncertainty of symmetry energy remain at suprasaturation densities, where the phase transitions of strong interaction matter and quark matter symmetry energy are likely to be taken into account. In this work, we have investigated the properties of symmetry energy by using a hybrid star with the hadron-quark phase transition. The quark matter interactions in hybrid stars are described based on 3-flavor NJL model with various vector and vector-isovector coupling constants. The ImMDI-ST interaction is then used to study the equation of state of nuclear matter by varying the parameters $x$, $y$, and $z$. In the present study, we found that all parameters and coupling constants of the interactions from the ImMDI-ST and NJL model can lead to widely different trends for the symmetry energy in the mixed phase and positions of the onset of the hadron-quark phase transition. The maximum mass of hybrid stars constrain mostly the EOS of symmetric matter (including symmetric nuclear and quark matter) at $2-5$ times saturation density. The different parameters $x$, $y$, and $z$ in the ImMDI-ST interactions will not lead to the variation of properties of symmetric nuclear matter, and thus have no significant effect on the maximum mass of hybrid stars. Compared to the vector-isovector interaction, the vector interaction of quark matter plays an more important role in mass-radius relations of hybrid stars. With increasing vector interaction coupling $R_V$ hybrid stars have larger maximum mass. This is due to that the vector interaction affects the EOS of not only asymmetric
quark matter but also symmetric quark matter. In addition, we also found that the radius and the tidal deformability of hybrid stars constrain mostly the density dependence of symmetry energy. Thus, the radius and formability of 1.4 $M_\odot$ are important constraints of the parameters $x$, $y$, and $z$ in ImMDI-ST interactions. However, the isospin properties of quark matter have no effect on the radius and formability of hybrid stars with a canonical mass 1.4 $M_\odot$, since there is not quark matter at densities from $\rho_0$ to $2\rho_0$.

So far, the following observables and constrains have been used for comparison: (1) the energy per nucleon for symmetric nuclear matter from the SCGF approach and the $\chi$EMBPT approach. (2) the nuclear symmetry energy $E_{sym}(\rho_0)$ and its slope parameter $L(\rho_0)$ from the analyses about nuclear experiments and astrophysical observations; (3)the observed maximum mass $2.08 \pm 0.07M_{\odot}$ for the two pulsars PSR J0740+6620; (4) the radius inferred from the X-ray bursts of LMXB $10.62 \leq R_{1.4} \leq 12.83$ km and the PSR J0740+6620 of NICER and XMM-Newton $R_{1.4}=12.45 \pm 0.65$ km; (5) the tidal deformability $70 \leq \Lambda \leq 580$ extracted by the LIGO and Virgo Collaborations. In addition, some of the new discoveries and observations provide more rigorous constraints on symmetric energy, or may also contain some new physics. For an example, the newly discovered compact binary merger GW190814 with a secondary component of mass $(2.50 \thicksim 2.67)M_{\odot}$, which can be reproduced by a super-fast pulsar~\cite{Zha20} or quark star~\cite{Zha21}. These constraints of massive compact stars can also be used to understand the properties of the hadron-quark phase transition. For example, the coupling constants $R_V$ and $R_{IV}$ in the NJL model determine the EOS matter and also affect the critical point as well as the QCD phase structure. To further explore the QCD phase structure and search for the signal of the critical point between the crossover and the first-order transition, experimental programs such as the beam-energy scan (BES) at RHIC and the compressed baryonic matter (CBM) at Facilities for Antiproton and Ion Research (FAIR) were proposed. The promising results are available to provide more constraints on the EOSs of quark and nuclear matter, which are helpful in the understanding of the QCD phase structure and isospin properties of compact stars.

\begin{acknowledgments}
This work is supported by the National Natural Science Foundation of China under Grants No. 11922514 and No. 11975132, and the Shandong Provincial Natural Science Foundation, China Grants No. ZR2021QA037 and No. ZR2019YQ01.
\end{acknowledgments}

\end{document}